\documentclass[dvips,twoside,fleqn]{article}
\usepackage{times}
\usepackage{amsmath}
\usepackage{amssymb}
\usepackage{amstext}
\usepackage{amssymb}
\usepackage{espcrc2}
\usepackage{graphicx}
\title{Abelian monopole condensation in lattice gauge theories}
\author{
P. Cea\address{Dipartimento di Fisica - Universit\`a di Bari - 
Via Amendola 173 -
70126 Bari - Italy}$^{\text{,b}}$,
and L. Cosmai\address{INFN - Sezione di Bari - Via Amendola 173 
- 70126 Bari - Italy}
}

\begin{document}

\begin{abstract}
We investigate the dynamics of lattice gauge theories in an 
Abelian monopole background field. By means of the gauge-invariant lattice
Schrodinger functional we study the Abelian monopole condensation in U(1) 
lattice gauge theory at zero temperature and in SU(3) lattice
gauge theory at finite temperature.
\end{abstract}
\maketitle

\section{Introduction}

The dual superconductivity mechanism
to explain color confinement has been suggested since the early day of 
QCD~\cite{dualsuperc}.
The first evidences for the dual superconductivity was obtained by studying
the dual Meissner effect~\cite{primistudi}.
More recently an alternative method to detect the dual superconductivity has been
proposed by the Pisa Group~\cite{DiGiacomo1995}: it consists in measuring 
a disorder parameter given in terms of an operator with non zero magnetic
charge and nonvanishing v.e.v. in the confined phase.
In the case of non Abelian Gauge theories they need
to perform the Abelian projection.
Indeed the Pisa Group found 
evidence of Abelian monopole condensation in several gauges:
plaquette gauge, butterfly gauge and Polyakov gauge~\cite{DiGiacomo1999}.

The aim of this work is 
to investigate the dynamics of lattice gauge theories in an
Abelian monopole background field in a gauge-invariant way.
We use the gauge-invariant effective action
for external background field  defined by means of the 
lattice Schr\"odinger functional~\cite{RossiTesta1980,Luscher1992}
\begin{equation}
\label{Zlatt}
{\mathcal{Z}} \left[ \vec{A}_a^{\text{ext}} \right] =
\int {\mathcal{D}}U \exp(-S_W) \,,
\end{equation}
where $S_{\text{W}}$ is the Wilson action and
$\vec{A}^{\text{ext}}(\vec{x}) =  \vec{A}_a^{\text{ext}}(\vec{x})
\lambda_a/2$ is the external field.
The integration constraint over the lattice links is
$U_\mu(x)|_{x_4=0} = U_\mu^{\text{ext}}(\vec{x},0)$,
where $U_\mu^{\text{ext}}$ is the lattice
version of the external field $A_\mu^{\text{ext}}$.
The Schr\"odinger functional is invariant under arbitrary lattice
gauge transformations of the boundary links.
The lattice effective action for the background field
$A_\mu^{\text{ext}}(\vec{x})$ is 
($L_4$ extension in Euclidean time):
\begin{equation}
\label{effact}
\Gamma\left[ \vec{A}^{\text{ext}} \right] = -\frac{1}{L_4}
\ln \left\{ \frac{{\mathcal{Z}}[U^{\text{ext}}]}{{\mathcal{Z}}[0]} \right\}
\end{equation}
${\mathcal{Z}}[0]$ is the  
lattice Schr\"odinger functional
without external background field  
(i.e. $U_\mu^{\text{ext}} =1$).
Note that due to the manifest gauge invariance of the
lattice background field effective action we do not need to fix the gauge.

\section{U(1)}

We are interested in the
effective action with a Dirac magnetic monopole background field.
In the continuum the Dirac magnetic monopole field with the Dirac string
in the direction $\vec{n}$ is:
\begin{equation}
\label{monop}
e \vec{b}({\vec{r}}) =  \frac{n_{\mathrm{mon}}}{2} 
\frac{ \vec{r} \times \vec{n}}{r(r - \vec{r}\cdot\vec{n})} \,.
\end{equation}
where, according to the 
Dirac quantization condition, 
$n_{\mathrm{mon}}$ is an integer
and $e$ 
is the electric charge 
(magnetic charge = $n_{\mathrm{mon}}/2e$).
We consider the gauge-invariant background field
action Eq.~(\ref{effact}) where the external
background field is given by the lattice  version of the Dirac 
magnetic monopole field.
In the 
numerical simulations
we put the lattice Dirac monopole at the center of the time slice $x_4=0$. 
To avoid the singularity due
to the Dirac string we locate the monopole between two neighbouring sites. 
We have checked that the results are not too sensitive to the 
precise position of the magnetic monopole.
We introduce the disorder parameter for confinement:
\begin{equation}
\label{mu}
\mu = e^{-E_{\mathrm{mon}} L_4} = 
\frac{{\mathcal{Z}}[{\mathrm{mon}}]}{{\mathcal{Z}}[0]}  \,,
\end{equation}
where ${\mathcal{Z}}[0]$
is the Schr\"odinger functional 
with $n_{\mathrm{mon}}=0$.
According to the physical interpretation of the effective action,
$E_{\mathrm{mon}}$ is the energy to create a monopole.
To avoid the problem of dealing with a partition function we consider
$E^\prime_{\mathrm{mon}} = \partial E_{\mathrm{mon}} / \partial \beta$,
that is analogous to the parameter 
$\rho$ introduced by the Pisa group~\cite{DiGiacomo1999}.
Note that $E^\prime_{\mathrm{mon}}$
is given by the difference between 
the average plaquette 
$P$ obtained from configurations without and with
the monopole field.

We performed lattice simulations on 
$16^4$, $24^4$ and $32^4$ lattices with 
periodic boundary conditions using the Quadrics/Q4 - QH1 in Bari. 
Note that the links belonging to the time slice $x_4=0$ 
and to the spatial boundary are constrained (no update).
The constraint on the 
links starting from sites belonging to the spatial boundary
corresponds in the continuum to the usual requirement that
the fluctuations over the
background field vanish at the spatial infinity.
The  contributions to $E^\prime_{\mathrm{mon}}$
due to the constrained links must be subtracted,
i.e.: only the dynamical links must be taken into account in evaluating
$E^\prime_{\mathrm{mon}}$. 
\begin{figure}[t]
\begin{center}
\includegraphics[width=0.45\textwidth,clip]{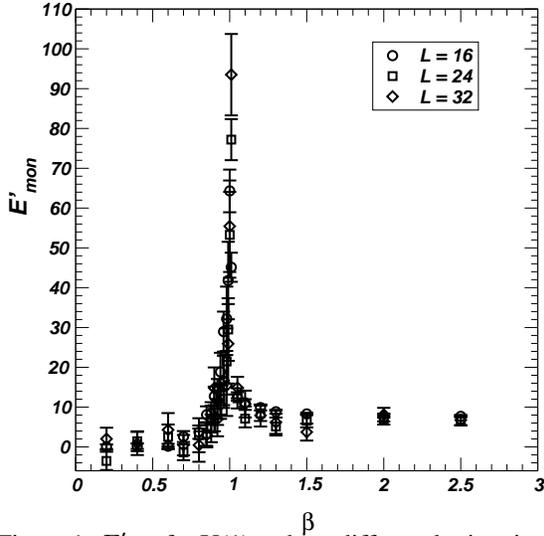}
\vspace{-1.24truecm}
\caption{$E^\prime_{\mathrm{mon}}$ for U(1) at three different lattice sizes.}
\end{center}
\end{figure}
In the strong coupling region
$\beta \lesssim 1$ the monopole energy is zero. 
This means that, according to Eq.~(\ref{mu}), 
the disorder parameter $\mu \simeq 1$.
Near the critical coupling $\beta_c \simeq 1$,
$E^\prime_{\mathrm{mon}}$
diplays a sharp peak which increases and shrinks by increasing the 
lattice volume. 
This means that the disorder parameter decreases towards zero in the 
thermodynamic limit when  $\beta \to \beta_c$. 
In the weak coupling region
($\beta \gg \beta_c$) 
the plateau in
$E^\prime_{\mathrm{mon}}$ indicates that  
the monopole energy tends to the classical monopole action which 
behaves linearly  in $\beta$.
In order to obtain $\mu$ 
we perform the numerical integration of
$E^\prime_{\mathrm{mon}}$:
\begin{equation}
\label{trapez}
E_{{\mathrm{mon}}}  = 
\int_0^\beta 
E^\prime_{\mathrm{mon}} \,d\beta^{\prime}
\end{equation}
We found that the disorder parameter $\mu$
is different from zero 
in the confined phase (i.e. the monopoles condense
in the vacuum). 
Moreover $\mu \to 0$ 
when $\beta \to \beta_c$ 
in the thermodynamic limit (the precise 
determination of $\beta_c$ require a F.S.S. analysis).
Our result is gauge-invariant for the manifest gauge 
invariance of the Schr\"odinger functional.

\section{SU(3)}

We have studied the Abelian monopole condensation
in pure SU(2) lattice gauge theory at finite temperature.
Here we restrict ourselves to the more interesting 
case of SU(3) gauge theory. 
In this case the maximal Abelian group is 
U(1) $\times$ U(1).
Therefore we have two different types of Abelian monopole. 
Let us consider, firstly, the
Abelian monopole field 
\begin{equation}
\label{monop1su3}
g \vec{b}^a({\vec{r}}) = \delta^{a,3} \frac{n_{\mathrm{mon}}}{2} 
\frac{ \vec{r} \times \vec{n}}{r(r - \vec{r}\cdot\vec{n})} \,.
\end{equation}
which we call the $T_3$-Abelian monopole.
Now the functional integration constraint amounts on the lattice
to fix the links belonging to the time slice $x_4=0$.
\begin{figure}[t]
\begin{center}
\includegraphics[width=0.45\textwidth,clip]{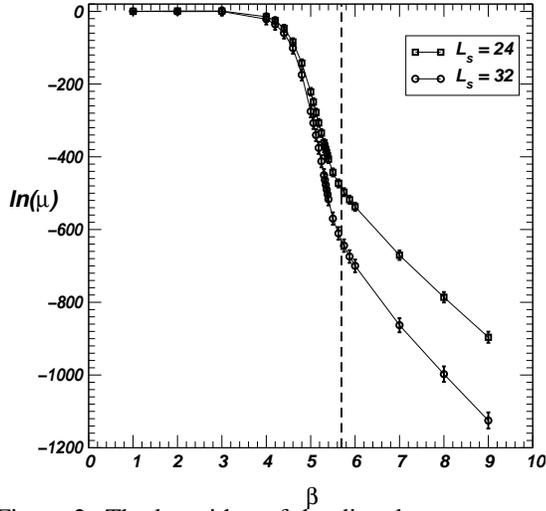}
\vspace{-1.24truecm}
\caption{The logarithm of the disorder parameter $\mu$ for 
the SU(3) ($T_3$ monopole) at two different spatial lattice sizes and
$L_t=4$. Tha dashed line is the critical coupling.}
\end{center}
\end{figure}
In the present case  the disorder parameter is defined as:
\begin{equation}
\label{mu1}
\mu = e^{-F_{\mathrm{mon}}/T} = 
\frac{{\mathcal{Z}}[{\mathrm{mon}}]}{{\mathcal{Z}}[0]}  \,,
\end{equation}
where $T=1/L_t$ is the temperature and 
$F_{\mathrm{mon}}$ is the free energy per monopole.
We measure
$F^\prime_{\mathrm{mon}} = \partial F_{\mathrm{mon}} / \partial \beta$.
Again this corresponds to measuring the 
difference between the average plaquette 
without and with the monopole field.

From Fig.~2 we see that in the thermodynamic limit the disorder parameter $\mu \sim 1$
in the confined phase, moreover 
$\mu \to 0$ when 
$\beta \to \beta_{\mathrm{c},\infty} = 5.6925(2)$~\cite{Iwasaki1992}
in the infinite volume limit.

The second type of Abelian monopole field is obtained from Eq.~(\ref{monop1su3}) 
replacing $\delta^{a,3}$ with $\delta^{a,8}$.
A previous study~\cite{DiGiacomo1999}
finds out that the disorder parameters for the two independent 
Abelian monopole defined by means of the Polyakov projection 
coincide within errors.
On the contrary, 
our numerical results show a 
dramatic difference for $F^\prime_{\mathrm{mon}}$. 
\begin{figure}[t]
\begin{center}
\includegraphics[width=0.45\textwidth,clip]{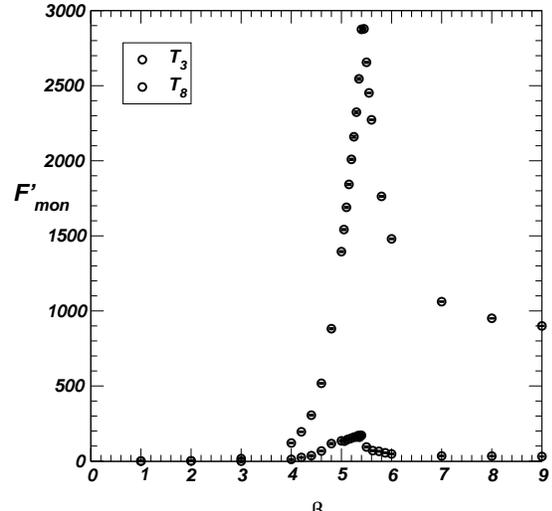}
\vspace{-1.24truecm}
\caption{$F^\prime_{\mathrm{mon}}$ for SU(3) for the 
$T_3$ and $T_8$ monopole on a $32^3 \times 4$ lattice.}
\end{center}
\end{figure}
The peak of $F^\prime_{\mathrm{mon}}$ 
in the case of the $T_8$-Abelian monopole is about
an order of magnitude greater
than in the $T_3$-Abelian monopole case (see Fig.~3). 
Consequently, in the former case, the disorder parameter
$\mu$ tends to zero more sharply.

\section{Conclusions}

We have studied the 
Abelian monopole condensation both in the Abelian gauge theory 
U(1) and finite
temperature non Abelian gauge theories SU(2) and SU(3).
We introduce a disorder parameter which signals 
the Abelian monopole condensation in the
confined phase. Our definition of the disorder parameter 
is by construction gauge
invariant. 
Our numerical results suggest that
the disorder parameter is different from zero in
 the confined phase and tends to zero when
the gauge coupling $\beta \to \beta_c$ in the thermodynamic
limit. Our estimate of the critical couplings are in
fair  agreement with the ones in the literarure. 
The precise determination of the
critical couplings and the critical exponents in the infinite volume 
limit could be obtained 
by means of the finite size scaling analysis.
In the case of the SU(3) gauge theory, 
there are two independent Abelian monopole
fields related to the two diagonal generators of the gauge group. 
Remarkably we find that the non perturbative vacuum reacts strongly in
the case of $T_8$-Abelian monopole. 
This seems to suggest that the vacuum
monopole condensate is predominantly formed by  $T_8$-Abelian monopoles.


\end{document}